\documentclass[twocolumn,showpacs,preprintnumbers,amsmath,amssymb,floatfix,superscriptaddress]{revtex4}

\usepackage{epsfig}
\usepackage{graphicx} 
\usepackage{dcolumn}  
\usepackage{bm}       
\usepackage{color}
\usepackage{epstopdf}

\usepackage{amsmath}
\usepackage{graphicx}

\newcommand{\be}{\begin{equation}}
\newcommand{\ee}{\end{equation}}

\usepackage{textcomp}

\begin{document}

\title{Ion induced density bubble in a strongly correlated one dimensional gas}

\author{J. Goold} \email{jgoold@phys.ucc.ie}

\affiliation{Department of Physics, University College Cork, Cork, Republic of Ireland} 
\affiliation{Centre for Quantum Technologies, National University of Singapore, Singapore}
\author{H. Doerk} 

\affiliation{Max-Planck-Institut f\"{u}r Plasmaphysik,
  Boltzmannstr. 3, D-85748 Garching, Germany}

 \author{ Z.~Idziaszek}

\affiliation{Institute of Theoretical Physics, Faculty of Physics, University of Warsaw, Ho\.{z}a 69, 00-681 Warsaw, Poland}

\author{T. Calarco}

\affiliation{Institute for Quantum Information Processing, University of Ulm, Albert-Einstein-Allee 11, D-89069 Ulm, Germany}

\author{Th. Busch} 

\affiliation{Department of Physics, University College Cork, Cork, Republic of Ireland}

\begin{abstract}
  We consider a harmonically trapped Tonks-Girardeau gas of
  impenetrable bosons in the presence of a single embedded ion, which is assumed to be tightly confined in a RF trap. In an ultracold ion-atom collision the ion's charge
  induces an electric dipole moment in the atoms which leads to an
  attractive $r^{-4}$ potential asymptotically. We treat the
  ion as a static deformation of the harmonic trap potential and model its
  short range interaction with the gas in the framework of quantum
  defect theory. The molecular bound states of the ionic potential are
  not populated due to the lack of any possible relaxation process in
  the Tonks-Girardeau regime. Armed with this knowledge we calculate
  the density profile of the gas in the presence of a central ionic
  impurity and show that a density \textit{bubble} of the order of a micron occurs around the ion for
  typical experimental parameters. From these exact results we show that an ionic impurity in a Tonks gas can be
  described using a pseudopotential, allowing for significantly
  easier treatment.
\end{abstract}
\maketitle

\textit{Introduction} - Since the days of superfluid helium the
introduction of charged particles to a superfluid has proven to be a
useful tool to explore fundamental physics. Statically they can
lead to fundamental structures such as the celebrated ion
\textit{snowballs} and electron \textit{bubbles} \cite{Benderskii:99},
whereas studying the motion of an immersed ion can be
used to characterise the superfluid nature of $^{4}$He
\cite{Rief:60}. Through the advances in the area of atom
trapping and laser cooling the number of known superfluids and the
environments in which they can be studied has in the last decade
vastly increased. Combining these degenerate fermionic and bosonic
gases, which come in a wide range of geometries and
dimensionalities \cite{Bloch:08} with cooled and trapped single ions
in Paul and Penning traps \cite{Wineland:03} is therefore
currently one of the most exciting and dynamic areas in physics.
Such research is motivated by our need to understand the true quantum
nature of matter, the quest for scalable quantum information
processing and also the possibility of efficient quantum simulation
\cite{qiprm:05}. However, despite the achievements in both the areas
of ion and atom physics, currently little is known about the
physics of an ion interacting with a many-body atomic system in the
ultracold regime \cite{Cote:02, Massignan:05}.

 By today most studies of atom-ion systems have focused on the scattering properties of single atoms and single
ions \cite{Dalgarno:00, Bodo:08, Julienne:09}. Controlled collisions
of a single atom and a single ion have been studied in
\cite{Zoller:07} with trap induced resonances predicted. Such
resonances along with the strong atom ion interaction have
  shown to be useful for enhancing the speed and the fidelity of a
collisional quantum gate \cite{Hauke:09}. On the many-body level, it
has been proposed to use an ion as a scanning tunnelling microscope
\cite{Kollath:07} which can in turn be used to measure the
energy distribution of, say, a Fermi gas in situ
\cite{Simmons:09}. Despite these promising proposals there is still
scarce knowledge of what is to be expected when an ion is introduced
to an ultracold many-body system. As the physics of a many-body
system is highly non-trivial even without the presence of an
ion, such studies are of paramount importance to shed light on the first results of the ongoing experimental progress
\cite{Grier:09, privcom:09, experiment}. Two situations have recently been
explored. In \cite{Cote:02} the authors consider an ion in a
homogeneous condensate and study the capture of atoms into the bound
states of the molecular potential. They predict the possibility
of observing mesoscopic molecular ions, where hundreds of the atoms can become bound in molecular states of the ion-atom potential. On the other hand, in \cite{Massignan:05} the authors consider a situation where the bound states are not accessible. They then estimate by using a
thermodynamical as well as a microscopic argument the number of
condensate atoms associated with an ion. Both studies suggest strong analogies with the snowball states observed in
the case of superfluid helium. In contrast to both of these studies,
in this work we demonstrate that in the case of a harmonically trapped
and strongly correlated one-dimensional gas, a micron sized
\textit{bubble} in the density may be observed around the ion.

\begin{figure}[tb]
  \begin{center}
    \includegraphics[width=\linewidth] {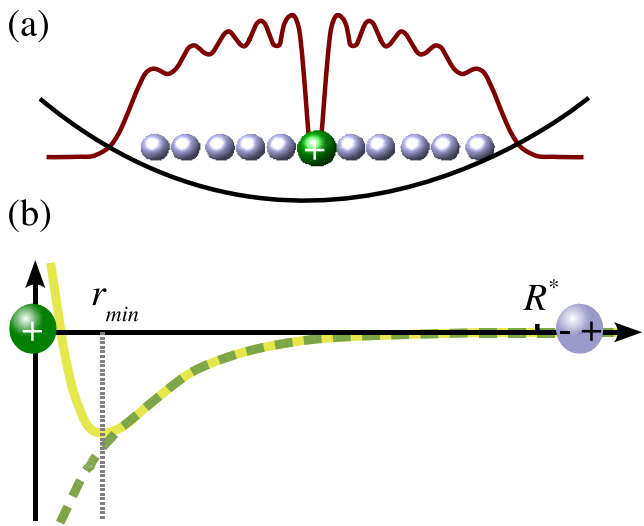}
  \end{center}
  \caption{Schematic showing (a) an positively charged ion
    embedded in a harmonically trapped Tonks-Girardeau gas and (b) the
    interaction potential for the ion-atom interaction under the
    Born-Oppenheimer approximation (dashed). The potential minimum is at $r_{min}$ and the polarisation
    length of the ion is indicated by $R^{*}$. The potential curve
    goes as $r^{-4}$ at large distances.}
  \label{fig:schematic}
\end{figure}

Strongly correlated one-dimensional quantum gases have recently
become available in ultracold physics labs worldwide. One of the
stand out achievements for cold atoms in low dimensions has been the
 first observation of the Tonks-Girardeau (TG) regime
\cite{Paredes:04, Kinoshita:04}. In this limit the gas is
defined to be a one-dimensional, strongly correlated gas \cite{Girardeau:60}. As it is exactly solvable
through the Bose-Fermi mapping theorem \cite{Girardeau:60} it is very
well suited to study its behaviour in the presence of impurities. The physics of impurities interacting with a
many-body system is of great importance in understanding many
complex condensed matter phenomena and as ultra-cold systems
allow to tailor the interaction they have proven to be versatile model systems for a large
number of different areas in physics \cite{Girardeau:09}. In fact, a recent experiment in Cambridge has
observed the motion of a spin impurity in a TG gas and reported interesting and complex dynamics \cite{Kohl:09}. In this letter we add
to the existing studies \cite{Cote:02, Massignan:05} of ionic
impurities interacting with atomic many-body systems by considering an
ion embedded in a TG gas (see Fig.~\ref{fig:schematic}(a)). Utilising
the Fermi-Bose mapping theorem \cite{Girardeau:60} we calculate the
single particle density of a TG gas in the presence of an ion. We
treat the ion's presence as a perturbation on the harmonic trap and
model its short range interaction in the framework of quantum defect
theory. 

\textit{Ultracold atom-ion interactions} - Under the Born-Oppenheimer
approximation, the interaction between an atom and a positively
charged ion, at large distance $r$ and low energies, is given by the
potential $V(r)=-\alpha e^{2}/2  r^{4}$. Here $\alpha$ is the
dipolar polarizability of the atomic species and $e$ is the charge on
the electron. This approximation is valid only if the atom is
in an atomic $s$ state as for higher electronic states the
Born-Oppenheimer potential significantly changes shape. A characteristic lengthscale of
this interaction can be found by equating the interaction energy
and the relative kinetic energy of an atom-ion pair. This gives
  the so-called polarisation length $R^{*}=\sqrt{{2\mu\alpha
e^{2}}/{\hbar^{2}}}$, where $\mu$ is the reduced mass and which allows to define the interaction energy of a typical atom ion
collision as $E^{*}={\hbar^{2}}/{2\mu (R^{*})^2}$.

The Hamiltonian for the axial dynamics of an atom and an ion in \textit{seperate} one dimensional harmonic traps maybe written as \cite{Zoller:07},
\begin{equation}
\label{effect_ham}
\mathcal{H}_{1D}=\sum_{\nu=i,a} \left(-\frac{\hbar^2}{2m_{\nu}}\frac{\partial^2}{\partial x_{\nu}^2} +\frac{1}{2}m_{\nu}\omega_{\nu}^{2}x^{2}_{\nu}\right) +V_{int}(|x_i-x_a|),
\end{equation}
where $V_{int}$ is the full one dimensional interaction potential, which at large distances has the same power law as in three dimensions i.e. $V(|x|)=-{\alpha
  e^2}/{2  x^{4}}$. In many realistic experimental scenarios the frequency of the ion trap will be many orders of magnitude greater than that of the atom trap and it is reasonable to approximate the ion to be located quasi-statically at the centre of the atom trap. In this approximation one may neglect the kinetic and potential energy of the ion in \eqref{effect_ham}. This allows us to write a single particle Hamiltonian which describes an atom in a harmonic trap which is deformed at the trap centre by the ionic interaction potential,
\begin{equation}
\label{eq:Hamiltonian_spl}
  \mathcal{H}=-\frac{\hbar^2}{2m_{a}}\frac{\partial^2}{\partial x_{a}^2} +\frac{1}{2}m\omega_{a}^2 x_{a}^{2} +V_{int}(x_{a}).
\end{equation}
The solution of the Schr\"{o}dinger equation for the single particle states is complicated by the fact that the real interaction potential $V_{int}$ begins to deviate from the long range asymptotic law and begins to quickly diverge towards $+\infty$ (see Fig.~\ref{fig:schematic}(b)). However, at this distance the effect of the harmonic trapping potential can safely be neglected and at short range we may approximate the scattering behaviour to be equivalent to a free atom-ion scattering event. The Schr\"{o}dinger equation, in this region, maybe written as 
\begin{equation}
  \label{eq:short range}
  \left(-\frac{\hbar^2}{2\mu}\frac{\partial^2}{\partial x^2} 
        -\frac{\alpha e^{2}}{2 x^{4}}\right)\psi_n(x)=\epsilon_{n}\psi_n(x),
\end{equation}
where $\mu$ is the reduced mass and $\phi_{e}$ and $\phi_{o}$ are the unknown scattering phases to be determined (see eqs. \eqref{eq:even} and \eqref{eq:odd} below ). In one dimension the solutions for eq.~\eqref{eq:short range} have even and odd parity. The asymptotic solutions, $|x|\rightarrow0$, to eq.~\eqref{eq:short range} are given by\cite{Zoller:07}
 \begin{align}
  \label{eq:even}
   \psi^{e} &\sim |x| \sin\left(\frac{R^*}{|x|}+\phi_e\right)\\
  \label{eq:odd}
   \psi^{o}&\sim\; x\; \sin\left(\frac{R^*}{|x|}+\phi_o\right)
\end{align}
The principle idea of the quantum defect theory is to replace the real interaction potential in eq.~\eqref{eq:Hamiltonian_spl}, $V_{int}$, with the asymptotic potential and use eqs. \eqref{eq:even} and \eqref{eq:odd} as boundary condition for the numerical solution of the Schr\"{o}dinger equation for Hamiltonian eq.~\eqref{eq:Hamiltonian_spl}. In analogy to the three-dimensional case the short range phase may be related to the one-dimensional even ($s-$wave) and odd ($p-$wave) scattering lengths of an ultracold atom-ion collision via $a_{1D}^{e,o}=\lim_{k\rightarrow 0}[\tan(\phi_{e,o}(k))/k]=-\cot(\phi_{e,o})$. Unfortunately the exact values for the scattering lengths of current experimental systems are not yet known and we therefore take $\phi_{e,o}$ to be adjustable parameters for now. The set of single particle states obtained using this approach are the ones which are used to calculate the groundstate of a Tonks-Girardeau gas in the presence of a centrally embedded ion.

\textit{The Tonks-Girardeau gas} - The effective one-dimensional
  Hamiltonian of a gas of $N$ bosons in trapping potential $V(x)$ can
  be written as $\mathcal{H}=\sum_{n=1}^N\left(-\frac{\hbar^2}{2m}\frac{\partial^2}{\partial x_n^2}+V\left(x_n\right)\right)+\sum_{i<j}V(|x_i-x_j|)$, where $m$ is the mass of a single atom. For low densities two
  body collisions dominate the dynamics, and for low temperature we
  can approximate these by employing a point-like potential of the
  form $V(|x_i-x_j|)=g_{1D}\delta(|x_i-x_j|)$. Here $g_{1D}$ is the
one dimensional coupling constant, which is related to the three
dimensional scattering length, $a_{3D}$, via $g_{1D}=\frac{4\hbar^2
  a_{3D}}{ml_\perp}\left(l_\perp-Ca_{3D}\right)^{-1}\;$, where $C$ is
a constant of value $C=1.4603\dots$ \cite{Olshanii:98} and $l_\perp$ is the transverse trapping length.
\begin{figure}[tb]
  \begin{center}
    \includegraphics[width=\linewidth] {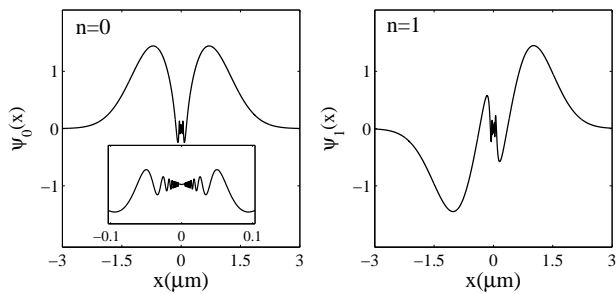}
  \end{center}
  \caption{The first two trap states of a harmonic trap with a
    centrally embedded ion for a $^{135}$Ba$^{+}$-$^{87}$Rb system for axial
    trapping frequency, $\omega=70Hz$. The inset for the $n=0$ state shows a zoom in of the region around
    zero where the wavefunction is rapidly oscillating. The short range phases are chosen to be $\phi_e=\frac{\pi}{4}$ and $\phi_o=-\frac{\pi}{4}$.}
  \label{fig:basisf}
\end{figure} 
In the Tonks-Girardeau limit of infinitely strong repulsion
$(g_{1D} \rightarrow \infty)$, the problem may be mapped
to the problem of non-interacting spinless fermions, which is the so
called Fermi Bose mapping theorem \cite{Girardeau:60}. The main idea
of the mapping is that one can treat the atom-atom interaction term by replacing it with the following boundary
condition on the allowed bosonic wave-function, $\Psi_B(x_1\dots
x_N)=0\quad \text{if} \quad |x_i-x_j|=0$ for $i\neq j$ and $1\leq
i\leq\ j\leq N$. As this is formally equivalent to the Pauli
exclusion principle, one can solve for the associated ideal fermionic
wave-function $\Psi_F(x_1,\dots,x_N)=\frac{1}{\sqrt{N!}}
\det_{n,j=1}^N [\psi_n(x_j)]$, and calculate the bosonic solution from
this by appropriate symmetrization, $\Psi_B = A(x_1,\dots,x_N)
\Psi_F(x_1,x_2,\dots,x_N)$, where the unit antisymmetric function is
given by $A=\prod_{1\leq i < j\leq N} \text{sgn}(x_i-x_j)$
\cite{Girardeau:60}. As a consequence of this mapping the single
particle density of a TG gas has the particularly simple form
\begin{align}
  \label{eq:spd}
  \rho(x)=\sum_{n=0}^{N-1}|\psi_{n}(x)|^{2}
\end{align}
and thus all one needs to know to solve the many particle problem are
the eigenstates of the single particle problem. 

\textit{Ion in a Tonks-Girardeau gas} - By combining the ideas of the
previous two sections we are now in a position to calculate the effect
of a centrally embedded ion on the ground state of a harmonically
trapped TG gas. The Hamiltonian of the combined system can be written
as
\begin{align}
  \label{eq:Hamiltoniantgi}
  \mathcal{H}=&\sum_{n=1}^N\left(-\frac{\hbar^2}{2m_{a}}\frac{\partial^2}{\partial x_{n}^2}+\frac{1}{2}m_{a}\omega_{a}^2 x_{n}^{2} +V_{int}(x_{n})\right)\nonumber\\
  &\quad+g_{1D} \sum^{N}_{i<j}\delta(x_i-x_j)\;.
\end{align}
We now focus on obtaining the eigenstates of the single particle part
of the Hamiltonian, i.e.~eq.~\eqref{eq:Hamiltonian_spl}. Using the
prescription of the Fermi-Bose mapping we neglect the interaction part of the Hamiltonian
and numerically solve the single particle Schr\"{o}dinger equation to
find the eigenstates $\psi_n$ and eigenvalues $\epsilon_n$. For this we use the iterative Numerov method \cite{Numerov:77} and impose the
eqs.~\eqref{eq:even} and \eqref{eq:odd} as short range boundary
conditions. 

Due to the infinitely diverging nature of the ionic potential, there
exists a spectrum of bound molecular states between the ion and the atoms \cite{Cote:02}. In order for these to be
populated, however, a relaxation process must be present that allows to take away the excess energy of the binding process. In an ultracold gases this process would be equivalent to a three body recombination. However for the TG gas the density-density
correlation function becomes suppressed on the length scale of the
interparticle distance \cite{Girardeau:01} - and thus no relaxation
process for capture is available. We therefore do not need to consider the
bound state spectrum in the following. 

\begin{figure}[tb]
    \begin{center}
      \includegraphics[width=\linewidth] {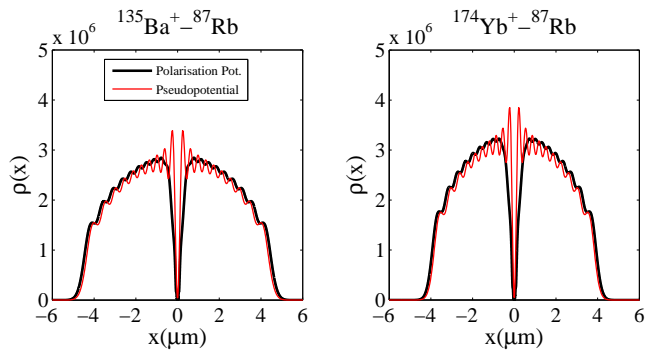}
      \label{fig:spds}
    \end{center}
    \caption{ The single particle densities of a TG gas of twenty
      particles in the presence of a central ion for
      $^{135}Ba^{+}$-$^{87}Rb$ and $^{174}Yb^{+}$-$^{87}Rb$ systems
      with a typical trapping frequency, $\omega=70$Hz (thick black
      line).  The short range phases in each case are chosen to be
      $\phi_e=\frac{\pi}{4}$ and $\phi_o=-\frac{\pi}{4}$. The thin red
      line in the plots represents the result of a pseudopotential
      approximation for the ion, using a large value for the scattering length.}
    \label{fig:spd}
\end{figure}

Our calculations are carried out for the experimentally relevant systems of $^{135}$Ba$^{+}$-$^{87}$Rb and
$^{174}$Yb$^{+}$-$^{87}$Rb. We assume a typical axial trap frequency of $\omega=70Hz$  for a one dimensional gas of $^{87}$Rb.
The $^{135}$Ba$^{+}$ and $^{174}$Yb$^{+}$ ions have typical polarisation lengths of $R^{*}=5544a_0$ and $6294a_{0}$, respectively, where $a_0$ is the Bohr radius.
For the $^{135}$Ba$^{+}$-$^{87}$Rb system we show the first two single particle eigenstates in
Fig.~\eqref{fig:basisf}. One can see that the centre of the
eigenstates is perturbed by the presence of the ion and a close up
of this central region is shown in the inset for the $n=0$
state. One can also see that in this region the wave function is rapidly
oscillating, with the exact details of the oscillations depending on
the short range phase. In the limit of such large trap lengths the
effects of the oscillations on the physics are only marginal and only
in the opposite limit to ours, if the traps were tight, do the
oscillations and hence the exact short range phase become
important. The fact that the anti-symmetric states are 
affected means that a certain amount of $p$-wave scattering is to be
  expected, however it will be very small in our wide trap.
 With the single particle wavefunctions at hand, we can now
  calculate the density distribution of a many-particle gas using
  eq.~\eqref{eq:spd}. In Fig.~\eqref{fig:spd} we show the density for
the two separate systems of $^{135}$Ba$^{+}$-$^{87}$Rb and
$^{174}$Yb$^{+}$-$^{87}$Rb. The thick black line shows the results using the quantum defect theory and
the Numerov method including the full atom-ion polarisation potential. It can be seen that the ion's
presence drastically perturbs the density distribution of the gas by
causing a bubble in the density around its position at $x=0$.
The size of this bubble is of the order of $1\mu m$. At first this
may seem counter-intuitive for an attractive impurity but one should recall
that scattering is repulsive at short distances (see Fig. \ref{fig:schematic}b) and with the lack of access to the bound states of the molecular potential in the TG regime - we see no build up of atomic density around the ion.

Let us finally note the possibility of treating atom-ion systems
  in the Tonks limit with a simpler approach, but approximating the
  exact interaction with a point-like potential \cite{Goold:08}. The
polarisation lengths for both $^{135}$Ba$^{+}$ and $^{174}$Yb$^{+}$ are, $0.29\mu m$ and $0.33\mu m$, and therefore smaller than trap lengths for one dimensional
systems, which for the traps considered here are $\sim1.3\mu m$. This suggests that a localised pseudopotential approximation can be used and in Fig.~\ref{fig:spd} we show this. In
Fig.~\ref{fig:spd} the pseudopotential approximation is shown by
the thin red line and one can see that the two descriptions
are in good agreement despite the fact that the scattering length in this case is not much smaller than the trap length. Therefore one can expect that the
pseudopotential approximation will encapsulate much of the essential
physics on the many-body level. In particular, the analytic
accessibility of the pseudopotential model allows one to easily calculate
the correlation functions and momentum distribution \cite{Pezer:07} of
a TG gas in the presence of a localised impurity \cite{Goold:08}. The pseudopotential model accurately describes the localised nature of the disturbance but due to it's zero width nature it underestimates the size of the bubble. It is also important to mention, where densities are concerned, that these results are also true for spin polarised samples of non-interacting fermions in one dimension.

\textit{Conclusions} - In this work we have shown that the
  presence of a single ion in a TG gas leads to the formation of a
  density bubble around the impurity. Using quantum defect theory we
have calculated the exact eigenstates of an atom-ion system in
one dimension and then applied the Bose-Fermi mapping theorem to
calculate the many-particle density. The bubble was found to be of the order of a
$\mu m$, and should therefore be observable in typical
experiments. We also outlined that the density can be very well
  approximated using a pseudopotential model under typical
  experimental constraints, which will allow to connect the atom-ion
  systems in the Tonks limit to a large number of results already
  existing in the literature.

{\it Acknowledgements} -JG and TB would like to thank Science
Foundation Ireland for support under contract 05/IN/I852 and
05/IN/I852STTF 08. JG would also like to thank J.~McCann, J.~Denschlag and T. Hennessy for important discussions. HDB and TC acknowlege support by the German Science Foundation through SFB TRR21 Co.Co.Mat and by the European Commission under the Integrated Project Scala.

\end{document}